\documentclass[twocolumn,%
secnumarabic,amssymb,amsmath,nobibnotes,aps,pre,showpacs,floatfix]{revtex4}
\usepackage{graphicx,hyperref}

\begin{document}

\title{Autocorrelations in the totally asymmetric simple exclusion process %
  and Nagel-Schreckenberg model}
\author{Jan de Gier}
\email{jdgier@unimelb.edu.au}
\affiliation{Department of Mathematics and Statistics,
  The University of Melbourne, VIC 3010, Australia} 
\author{Timothy M. Garoni}
\email{t.garoni@ms.unimelb.edu.au} 
\affiliation{ARC Centre of Excellence for Mathematics and Statistics of Complex
  Systems,
  Department of Mathematics and Statistics, The University of Melbourne,
  VIC 3010, Australia} 
\author{Zongzheng Zhou}
\affiliation{
Hefei National Laboratory for Physical Sciences at Microscale and 
Department of Modern Physics,
University of Science and Technology of China, 
Hefei, Anhui 230026, China}
\date{July 7, 2010}
\begin{abstract} We study via Monte Carlo simulation the dynamics of
  the Nagel-Schreckenberg model on a finite system of length $L$ with
  open boundary conditions and parallel updates.  We find numerically
  that in both the high and low density regimes the autocorrelation
  function of the system density behaves like $1-|t|/\tau$ with a finite
  support $[-\tau,\tau]$.  This is in contrast to the usual exponential
  decay typical of equilibrium systems.  Furthermore, our results
  suggest that in fact $\tau=L/c$, and in the special case of maximum
  velocity $v_{\rm max}=1$ (corresponding to the totally asymmetric
  simple exclusion process) we can identify the exact dependence of $c$
  on the input, output and hopping rates.  We also emphasize that the
  parameter $\tau$ corresponds to the integrated autocorrelation time,
  which plays a fundamental role in  quantifying the statistical errors
  in Monte Carlo simulations of these models.
\end{abstract}

\pacs{05.40-a, 05.60cd, 05.70Ln}
\maketitle

\section{Introduction}
\label{introduction}
The totally asymmetric simple exclusion process
(TASEP)~\cite{Spitzer70} is a simple transport model, of fundamental
importance in nonequilibrium statistical mechanics.  In addition to
its mathematical richness, it has applications ranging from molecular
biology to freeway traffic.

\begin{figure}[b]
  \caption{\label{tasep rules} A TASEP with $L=10$.
  }
  \includegraphics[scale=1]{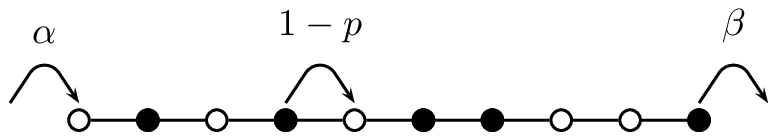}
\end{figure}

A TASEP consists of a chain of length $L$, with each site being either
occupied by a particle or not, on which particles hop from left to
right.  See Fig.~\ref{tasep rules}.  If site $i=1$ is vacant a
particle will enter the system with probability $\alpha$.  If site
$i=L$ is occupied the particle will leave the system with probability
$\beta$.  In the bulk of the system, a particle on site $i$ will hop
to site $i+1$ with probably $1-p$ provided $i+1$ is vacant, otherwise
it remains at site $i$.

TASEPs exhibit boundary-induced phase transitions, governed by the
parameters $\alpha$, $\beta$ and $p$.  In general, for a given $p$,
there exist three possible phases, depending on $\alpha$ and $\beta$:
a low-density phase, a high-density phase, and a maximum-current (or
maximum-flow) phase.

In the context of traffic models, it is most appropriate to update all
sites in parallel at each time-step.  The stationary distribution of
the TASEP with fully-parallel
updates~\cite{deGierNienhuis99,EvansRajewskySpeer99} is known exactly.
(For reviews of the stationary properties of TASEPs with random
sequential updates see~\cite{Derrida98,Schutz01}.)  The
Nagel-Schreckenberg (NaSch) model~\cite{NagelSchreckenberg92} is an
important generalization of the parallel-update TASEP, in which
particles can move up to $v_{\max}\in\mathbb{N}$  sites per time
step. The NaSch model is generally considered to be the minimal model
for traffic on freeways~\cite{ChowdhurySantenSchadschneider00}.  
While many results are known rigorously for
the TASEP, our understanding of the NaSch model and its further
generalizations typically rely on numerical simulation.  This is
particularly true of traffic network models, in which the NaSch model
is often a component (see for example
\cite{EsserSchreckenberg97,SchreckenbergNeubertWahle01,%
CetinNagelRaneyVoellmy02}).

In the current article we focus on dynamic (auto)correlation
functions. The autocorrelations of the TASEP with random sequential
update have been studied in
\cite{PierobonParmeggianiVonOppenFrey05,AdamsZiaSchmittmann07} and
display a separation of time scales between relaxation of local
density fluctuations and collective domain wall motion. In particular,
it was recently observed~\cite{AdamsZiaSchmittmann07} that the TASEP
with random sequential update exhibits non-trivial oscillations in the
power spectrum of the system density, in the low and high density
phases. In this article, we further elucidate the nature of these
non-trivial oscillations, and demonstrate that they extend to the
NaSch model generally. We emphasize that all the simulations performed 
in this work used fully-parallel updates, including our simulations of
TASEP (which we view as the special case of the NaSch model with 
$v_{\max}=1$).

\subsection{Density autocorrelations}
The system density, $n$, which is simply the fraction of sites which
are occupied, is an important quantity  in many applications,
including traffic modeling. The relationship between density and flow
is known as the {\em fundamental diagram} in the traffic engineering
literature.  While the stationary-state expectation $\langle n \rangle$ of $n$
is well understood for the NaSch model, 
and in fact known rigorously for the TASEP,
the dynamic behavior of $n_t$ is non-trivial.
In this article we numerically study the autocorrelation function
$\rho_n(t):=(\langle n_0 n_t\rangle - \langle n\rangle^2)/\text{var}(n)$
of the general NaSch model,
and find a very simple form for its finite-size scaling.
Up to very small corrections, our simulations show that in
both the high and low density phases we simply have
\begin{equation}
  \rho_n(t) =
  \begin{cases}
    1 - |t|/\tau, & |t| \le   \tau, \\
    0,            & |t| \ge \tau, \\
  \end{cases}
\label{rho conjecture}
\end{equation} for some constant $\tau\propto L$.

The linear decay in \eqref{rho conjecture} is in sharp contrast to the 
usual exponential decay typical of equilibrium systems. 
In fact, as discussed in section~\ref{infinite support section}, there
are good theoretical reasons to believe that $\rho_n(t)$ must ultimately 
decay exponentially on sufficiently long time scales, rather than exhibit
the strictly finite support suggested by \eqref{rho conjecture}.
However, as demonstrated by the simulations in sections~\ref{ASEP} 
and~\ref{NaSch}, any corrections to the finite-support behavior 
displayed in \eqref{rho conjecture} are extremely weak, and in practice
\eqref{rho conjecture} provides a very accurate approximation to the 
behavior of $\rho_n(t)$ throughout the low and high density phases.
In particular, \eqref{rho conjecture} provides a very good
approximation to $\rho_n(t)$  for values of $p$ relevant for traffic
modeling. 

The Fourier series of $\rho_n(t)$ gives the power spectrum of $n$, and we note
that taking the Fourier series of \eqref{rho conjecture} does indeed
produce oscillations as reported in~\cite{AdamsZiaSchmittmann07}. 
Indeed, we have
\begin{equation}
\sum_{t=-\infty}^{\infty} \rho_n(t)\, e^{i\,\omega\,t} 
= \frac{1}{\tau} \frac{1-\cos \tau\,\omega} {1-\cos \omega}.
\label{spectral density conjecture}
\end{equation} 
The discussion in~\cite{AdamsZiaSchmittmann07} focused
on the case $v_{\max}=1$, with random sequential updates.
However, our simulations show that \eqref{rho conjecture}, and hence 
\eqref{spectral density conjecture}, hold more generally for the NaSch model
with arbitrary $v_{\max}$.

The specific form \eqref{rho conjecture} of the autocorrelation
function has some interesting consequences for the design of Monte
Carlo simulations. In particular, 
as discussed in section~\ref{autocorrelations},
assuming the validity of \eqref{rho conjecture} we immediately have 
$\tau=2\,\tau_{\text{int},n}$ where $\tau_{\text{int},n}$ is the 
{\em integrated autocorrelation time} of $n$.  
The integrated autocorrelation time can be interpreted loosely
as the number of time steps between ``effectively independent''
samples.  It is therefore reasonable to conjecture that the parameter
$\tau$ should equal the amount of time it takes a fluctuation of the
stationary state to traverse the system.  If we let $v$ denote the
speed of such a fluctuation then we might reasonably expect that
$\tau=L/v$.  In section~\ref{ASEP} we present numerical results that
strongly suggest that in fact, for TASEP, we have
\begin{equation}
\tau = L/|v_c(\alpha,\beta,p)|
\label{tau conjecture}
\end{equation}
where $v_c(\alpha,\beta,p)$, the {\em collective velocity}%
~\cite{KolomeiskySchutzKolomeiskyStraley98,deGierNienhuis99},
is known exactly. 
The results \eqref{rho conjecture} and \eqref{tau conjecture}  are
consistent with the suggestions in~\cite{AdamsZiaSchmittmann07} that
the physical origins of the observed oscillations in the power
spectrum of $n$ are related to the time needed for a fluctuation to
traverse the entire system.

Furthermore, while no exact expression for
$v_c(\alpha,\beta,p,v_{\max})$ is known for the general NaSch
model, 
the simulations presented in section~\ref{NaSch} demonstrate that
the scaling form \eqref{tau conjecture} extends to general $v_{\max}$.
In addition, in the deterministic limit ($p=0$)
simple physical arguments produce an exact relationship between 
$v_c$ and $v_{\max}$ which is in excellent agreement with the numerical 
results.

The remainder of this article is organized as follows. In
section~\ref{autocorrelations}, we briefly review some pertinent
general theory relating to autocorrelations and then discuss some general 
consequences of \eqref{rho conjecture}.
In section~\ref{ASEP}, we present our numerical evidence supporting
\eqref{rho conjecture} and \eqref{tau conjecture} for TASEP, and also
describe the exact expression for $v_c(\alpha,\beta,p)$ in this case.
We also explain relationship between \eqref{rho conjecture} and 
\eqref{tau conjecture} and the results presented 
in~\cite{AdamsZiaSchmittmann07}.
In section~\ref{NaSch}, we briefly review the definition of the NaSch
model before presenting our numerical results for $\rho_n(t)$ in this
case. Finally, we conclude in section~\ref{discussion} with a discussion.

\section{Autocorrelations}
\label{autocorrelations}
We begin by briefly recalling some standard definitions and results.
Consider a Monte Carlo simulation of an ergodic Markov chain, and
assume that sufficient time has passed that the system has reached
stationarity. If one now measures an observable $X$ at each time step
one obtains a stationary time series $X_1,X_2,\dots$ whose
autocovariance function is defined to be
\begin{equation}
  C_X(t) := \langle X_0 X_t\rangle - \langle X_0\rangle^2.
\end{equation}
The expectation $\langle\cdot\rangle$ here is with respect to the
stationary distribution, and we note that $C_X(0)=\text{var}(X)$.
The corresponding autocorrelation function is then defined as
\begin{equation}
  \rho_X(t) := \frac{C_X(t)}{C_X(0)}.
\end{equation}
Finally, assuming $C_X(t)$ to be absolutely summable,
its Fourier transform defines the spectral density
\begin{equation}
f_{X}(\omega) := \sum_{t=-\infty}^{\infty}C_X(t)\,e^{i\,\omega\,t}.
\label{spectral density}
\end{equation}

The spectral density is closely related to the Fourier transform of the 
time series. Specifically, given any stationary time series $X_1,X_2,\ldots,X_T$
we can define its discrete Fourier transform to be
\begin{equation}
\widehat{X}(\omega) := \frac{1}{\sqrt{T}}\sum_{t=1}^T X_t \,e^{i \omega t}
\label{discrete fourier transform}
\end{equation}
with $\omega=2 \pi m/T$ and $m=0,1,\ldots,T-1$.
It is then straightforward to show~\cite{ShumwayStoffer06} that for large $T$ 
we have
\begin{equation}
  \left\langle\big|\widehat{X}(\omega)\big|^2\right\rangle 
  = f_X(\omega) + O\left(\frac{1}{T}\right).
  \label{periodogram as spectral density}
\end{equation}

\subsection{Autocorrelation times}
\label{autocorrelation times section}
We now discuss the implications of the general form 
\eqref{rho conjecture} on two key time scales,
the {\em integrated} autocorrelation time and the {\em exponential} 
autocorrelation time.

\subsubsection{Integrated Autocorrelation time}
\label{integrated autocorrelation times section}
From $\rho_{X}(t)$ the integrated autocorrelation time is 
defined~\cite{SokalLectures} as
\begin{equation}
\tau_{{\text{int}},X}:=\frac{1}{2}\sum_{t=-\infty}^{\infty}\,\rho_{X}(t).
\label{tauint definition}
\end{equation}
If $\overline{X}$ denotes the sample mean of $X_1,X_2,\dots X_T$ then
the variance of $\overline{X}$ satisfies~\cite{SokalLectures}
\begin{equation}
  \text{var}(\overline{X}) 
  \sim 2\,\tau_{\text{int},X} \,\frac{\text{var}(X)}{T}, \qquad T\to\infty. 
  \label{tau int in error bars}
\end{equation}
It is \eqref{tau int in error bars} that accounts for the key role
played by the integrated autocorrelation time in the statistical
analysis of Markov-chain Monte Carlo time series.  If instead of a
correlated time series, one considers a sequence of independent random
variables, then the variance of the sample mean is simply
$\text{var}(X)/T$.  It is in this sense that $\tau_{\text{int},X}$
determines how many time steps we need to wait between two
``effectively independent'' samples.

It can now be seen immediately from \eqref{tauint definition} that, 
as noted in the introduction, \eqref{rho conjecture} and \eqref{tau conjecture}
imply
\begin{align}
\label{tauintASEP}
    2\tau_{\text{int},n} 
    &= \sum_{t=-\lfloor \tau\rfloor}^{\lfloor \tau\rfloor} 
    \left(1 - \frac{|t|}{\tau}\right),\\
    &= \tau + O(\tau^{-1})\\
    &= \frac{L}{|v_c|} + O(L^{-1}).
    \label{tau int with discrete corrections}
\end{align}
Equation \eqref{tau int with discrete corrections} provides a very simple
exact expression for $\tau_{{\rm int},n}$ in terms of the physical
parameters of the model. It is quite rare to have such an expression
for a non-trivial model.

\subsubsection{Exponential Autocorrelation time}
\label{Exponential autocorrelation times section}
Typically, we expect that $\rho_X(t)\sim \exp(-t/\tau_{\exp})$ as 
$t~\to~\infty$, which defines the exponential autocorrelation time 
$\tau_{\exp}$.
More precisely~\cite{SokalLectures}, one defines
the exponential autocorrelation time of observable $X$ to be
\begin{equation}
\tau_{{\text{exp}},X}:=\limsup_{|t|\to\infty}
\frac{-|t|}{\log\,\rho_{X}(t)},
\label{tauexp definition}
\end{equation}
and then the exponential autocorrelation time of the system as
\begin{equation}
\tau_{\text{exp}}:=\sup_{X}\tau_{{\text{exp}},X},
\end{equation}
where the supremum is taken over all observables $X$. The
autocorrelation time $\tau_{\exp}$ measures the decay rate of the
slowest mode of the system, and it therefore sets the scale for the
number of initial time steps to discard from a simulation, in order to
avoid bias from initial non-stationarity.
All observables that are not orthogonal
to this slowest mode satisfy $\tau_{{\text{exp}},X}=\tau_{\text{exp}}$.

For the TASEP in continuous time, $\tau_{\rm exp}$ was computed analytically
in~\cite{deGierEssler05,deGierEssler06} using the exact Bethe Ansatz
solution. In particular, it was found that $\tau_{\exp}$ is $O(1)$
with respect to $L$ in the high and low density phases. 
We would expect the same behavior to hold generally for the NaSch model.

However, if $\rho_n(t)$ were to have strictly finite support as claimed in
\eqref{rho conjecture}, then we would have $-|t|/\log\rho_n(t)=0$
for all $|t|>\tau$, implying that $\tau_{\exp,n}\neq \tau_{\exp}$. 
This would then mean that $n$ is orthogonal to the slowest relaxation mode, 
which seems implausible. We thus conclude that although \eqref{rho conjecture}
provides a very good approximation, $\rho_n(t)$ cannot actually have a strictly
finite support.

\subsection{\texorpdfstring{Finite-size scaling of $\rho_n(t)$}%
  {Finite-size scaling of density autocorrelation function}}
\label{infinite support section}
To obtain a more precise ansatz for $\rho_n(t)$ we therefore fix some
$k\in \mathbb{N}$ satisfying $k\le\lfloor\tau\rfloor$ and set
\begin{equation}
\rho_n(t)=
\begin{cases}
  1 - |t|/\tau, & |t| \le k,\\
  B\,e^{-|t|/\tau_{\exp}}, & |t| \ge k +1.\\
\end{cases}
\end{equation}
Since we know empirically that \eqref{rho conjecture} is a very good
approximation, it must be the case that $k/\tau\sim1$ as
$\tau\to\infty$.  Let us then write $\tau = k + \varepsilon$, where
the only assumption we make regarding $\varepsilon$ is that
$\varepsilon/\tau\to 0$ as $\tau\to\infty$.  Since the continuum limit of
$\rho(x\,\tau)$ should define a continuous function of
$x\in\mathbb{R}$ we choose the parameter $B$ by demanding that
$1-|t|/\tau=Be^{-|t|\tau_{\exp}}$ when $|t|=k$, which yields
\begin{equation}
  \rho_n(t) =
  \begin{cases}
    1 - |t|/\tau,                               & |t|\le k,\\
    \varepsilon\,e^{-(|t|-k)/\tau_{\exp}}/\tau, & |t|\ge k.\\
  \end{cases}
  \label{infinite support ansatz}
\end{equation}

It is worth noting that the two expressions \eqref{rho conjecture} and 
\eqref{infinite support ansatz} lead to the same leading-order 
expression \eqref{tau int with discrete corrections} for $\tau_{{\rm int},n}$.
Indeed, inserting \eqref{infinite support ansatz} into \eqref{tauint definition}
we obtain
\begin{equation}
2\tau_{\text{int},n}
=
\tau + \left(\varepsilon(1-\varepsilon) + 
\frac{2\varepsilon}{e^{1/\tau_{\exp}}-1}\right)\frac{1}{\tau}.
\end{equation}
Since $(e^{1/\tau_{\exp}}-1)^{-1} = O(1)$ for $\tau_{\exp}=O(1)$, 
the terms arising from the exponential decay of $\rho_n(t)$
are $O(\varepsilon)$ in the low and high density phases.

\section{TASEP}
\label{ASEP}
We begin this section by comparing the power spectrum found 
in~\cite{AdamsZiaSchmittmann07} with the Fourier transform of 
\eqref{rho conjecture}. We then present the exact result for the
collective velocity for TASEP, before presenting the results 
of our simulations.

\subsection{Power spectrum}
Let $N$ denote the number of occupied sites in a TASEP system, and
let $\widehat{N}(\omega)$ denote the discrete Fourier transform of a
particular time series $N_1,N_2,\ldots,N_T$, as defined 
in~\eqref{discrete fourier transform}.
The quantity $I(\omega):=T\langle\big|\widehat{N}(\omega)\big|^2\rangle$ is what
\cite{AdamsZiaSchmittmann07} refer to as the power spectrum of $N$.
They find that for the continuous-time TASEP in the low-density phase
\begin{equation}
\frac{I(\omega)}{T} \approx \frac{2\,v}{\omega^2}\frac{A}{D} 
\left[1-e^{-D\omega^2L/v^3}\cos\left(\frac{L\,\omega}{v}\right)\right],
\label{powerspec_continuous}
\end{equation}
where $A,D$ and $v$ are parameters, which \cite{AdamsZiaSchmittmann07}
set empirically to $v\approx0.4$, $D\approx20$ and $A\approx1/500$.

We now attempt to compare \eqref{powerspec_continuous} with the
corresponding result derived from \eqref{rho conjecture}. From
\eqref{periodogram as spectral density} we see that 
$f_N(\omega)\sim \langle\big|\widehat{N}(\omega)\big|^2\rangle$ 
as $T\to\infty$,  hence we should compare \eqref{powerspec_continuous} with
$f_N(\omega)=L^2\,f_n(\omega)$, where $f_n(\omega)$ is computed via
\eqref{rho conjecture}.  Although our empirical observations of the
behavior \eqref{rho conjecture} were made in the discrete time case of
fully-parallel updates,  \eqref{rho conjecture} can be interpreted as
a well defined continuous function on $\mathbb{R}$. 
In fact, the fully-parallel update rule becomes equivalent to the
random sequential update in the limit $\varepsilon\rightarrow 0$ of 
rescaled variables $1-p=\varepsilon$, $\alpha=\tilde{\alpha}\varepsilon$ 
and $\beta=\tilde{\beta}\varepsilon$. Here, $\tilde{\alpha}$ and $\tilde{\beta}$
are the usual injection and extraction rates of the TASEP in continuous time.

To compare with the continuous time result \eqref{powerspec_continuous}, we
compute $f_n(\omega)$ via the continuous-time Fourier transform, so
that \eqref{rho conjecture} and \eqref{tau conjecture} predict
\begin{equation}
  f_N(\omega) = \frac{2\,|v_c|}{\omega^2}\,\text{var}(n)\,L\,
  \left[1-\cos\left(\frac{L\,\omega}{|v_c|}\right)\right].
  \label{spectral density final conjecture}
\end{equation}
Now, since $\omega=2\pi m/T$, for sufficiently large $T$ we have
$\exp(D\omega^2 L/v^3)\approx 1$.  This is exactly the regime used
by~\cite{AdamsZiaSchmittmann07} in their Fig.~3 ($L=1000$ or $L=32000$
and $T=10^6$).  Therefore, in this regime we can identify
\eqref{powerspec_continuous} with \eqref{spectral density final conjecture}
if $v=|v_c|$ and
\begin{equation}
\frac{A}{D} = \text{var}(n)\,L.
\label{parameter matching}
\end{equation}

Some remarks are in order. Firstly, for the deterministic ($p=0$) 
parallel-update TASEP,
the static variance $\text{var}(n)$ can be computed analytically from
the known results for the two-point function \cite{EvansRajewskySpeer99}.
In the low density phase it is given by
\begin{equation}
\text{var}(n) = \frac{\alpha(1-\alpha)}{(1+\alpha)^3}\frac1L + O(L^{-2}),
\end{equation}
and for the high density region $\alpha$ is replaced by $\beta$. We
expect that $\text{var}(n)=O(1/L)$ would remain true when $p>0$,
and indeed for $v_{\max}>1$ as well. In general, therefore, we expect the 
prefactor in \eqref{spectral density final conjecture} to be $O(1)$ in $L$.

Finally, we note that \cite{AdamsZiaSchmittmann07} fit
\eqref{powerspec_continuous} to their data with a very small value of
$A/D$.  This small value follows from the fact that the numerical
simulations in \cite{AdamsZiaSchmittmann07} were performed along the
mean field line of the TASEP with random sequential update,  where,
theoretically, $\text{var}(n)$ is identically zero.  It is surprising
that \cite{AdamsZiaSchmittmann07} were still able to extract a
meaningful signal on this line. 

\subsection{Collective velocity}
The stationary distribution of the TASEP with fully-parallel
updates~\cite{deGierNienhuis99,EvansRajewskySpeer99} is known exactly.
In particular, if $\alpha<\beta, 1-\sqrt{p}$ such TASEPs reside in a
low-density phase, while for $\beta<\alpha,1-\sqrt{p}$ a high-density
phase results, with $\alpha = \beta < 1-\sqrt{p}$ defining a
coexistence line of the two phases (corresponding to a first order
phase transition). For $\alpha,\beta>1-\sqrt{p}$ by contrast, the
system resides in a maximum-current phase, in which the density is
precisely $1/2$.

The {\em collective velocity}~\cite{KolomeiskySchutzKolomeiskyStraley98}
is the drift of the center of mass of a momentary local fluctuation of
the stationary state,
and is related to the current (flow) $J$ and bulk density $\rho_b$ via
$v_c  = \partial J(\rho_b)/\partial \rho_b$.
An exact expression for $v_c(\alpha,\beta,p)$ is 
available~\cite{deGierNienhuis99} for the case of parallel-update TASEP.
If we define, for convenience, the function
\begin{equation}
g(x,p) = \frac{(1-p)((1-x)^2-p)}{(1-x)^2 + p(2x-1)},
\end{equation}
then 
\begin{equation}
v_c(\alpha,\beta,p)
=
\begin{cases}
 g(\alpha,p), & \text{low density phase},\\
-g(\beta,p),  & \text{high density phase}.
\label{TASEP collective velocity}
\end{cases}
\end{equation}
The negativity of the collective
velocity in the high-density phase is simply due to the fact that it is the
propagation of holes from right to left, rather than of particles from
left to right, that is important in this phase.

Using these exact expressions for $v_c$ the expression~\eqref{tau conjecture}
now becomes
\begin{equation}
  \tau =
  \begin{cases}
    L/g(\alpha,p), & \alpha < \beta, 1 - \sqrt{p},\\
    L/g(\beta,p),  & \beta  < \alpha, 1 - \sqrt{p}.\\
  \end{cases}
  \label{ASEP tau int conjecture}
\end{equation}
We note that for $p=0$ we have $|v_c|=1$ identically throughout the
high and low density regimes so that we simply have $\tau=L$ in this
case.  We also note that in the low-density (high-density) phase
$\tau$ is independent of $\beta$ ($\alpha$).

\subsection{Simulations}
We now turn our attention to our Monte Carlo simulations.  We
simulated the parallel-update TASEP at a variety of values of
$\alpha,\beta$ and $p$ corresponding to both the low and high density
phases, for system sizes $L=10^3$, $5\times 10^3$ and $10^4$.  Each
simulation consisted of $10^4 L/v_c$ iterations, with the first 
$10^3 L/v_c$ time-steps discarded to ensure negligible bias due to initial
non-stationarity (initially the system was empty).  Assuming the
validity of~\eqref{tau conjecture}, this implies we generated
$1.8\times 10^4\,\tau_{\text{int},n}$ samples of the stationary
distribution in each simulation. 

\begin{figure}[ht]
  \includegraphics[scale=0.6]{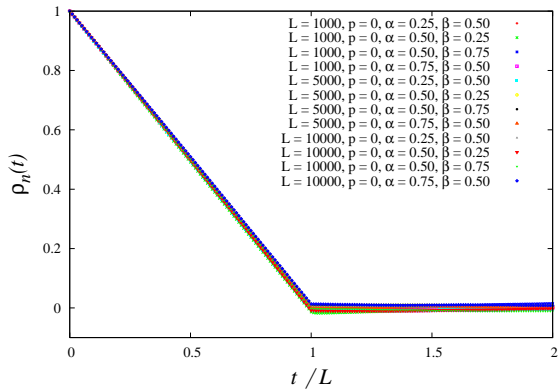}
  \caption{\label{asep LDHD p=0} Color online.
    Finite-size scaling plot of $\rho_n(t)$ for $p=0$ parallel-update TASEP in 
    the high-density and low-density phases, 
    for $L=10^3, 5\times10^3,10^4$ and a variety of $\alpha,\beta$.
  }
\end{figure}

For each simulation, we measured $n$ at each iteration, and from the resulting
time series we estimated the autocorrelation function $\rho_n(t)$
using the standard estimators~\cite{SokalLectures}. Fig.~\ref{asep LDHD p=0}
shows a finite-size scaling plot of $\rho_n(t)$ assuming the ansatz
given by \eqref{rho conjecture} with $\tau=L$, in the $p=0$ case.  The
agreement is clearly very good, and the sharpness of the cusp at $t=L$
suggests that any corrections to the finite-support ansatz 
\eqref{rho conjecture} are very small.

\begin{figure}[ht]
  \includegraphics[scale=0.6]{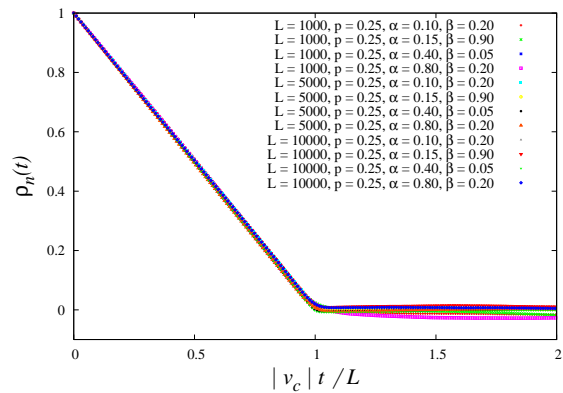}
  \caption{\label{asep LDHD p=0.25} Color online.
    Finite-size scaling plot of $\rho_n(t)$ for $p=0.25$ parallel-update TASEP 
    in the high-density and low-density phases, 
    for $L=10^3, 5\times10^3,10^4$ and a variety of $\alpha,\beta$.
    The choices of $\alpha,\beta$ shown correspond to four distinct values of 
    $v_c$ providing strong evidence for the conjecture 
    \eqref{ASEP tau int conjecture}.}
\end{figure}

\begin{figure}[ht]
  \includegraphics[scale=0.6]{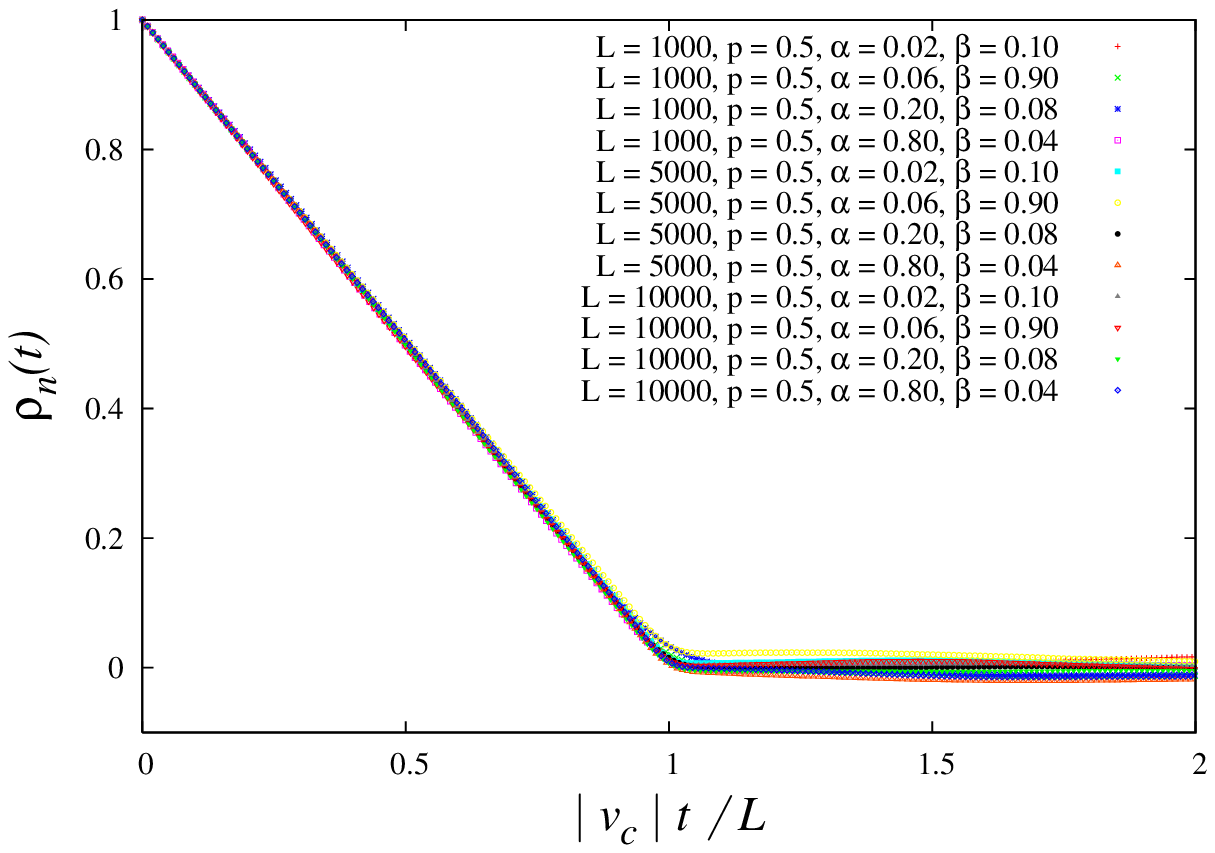}
  \caption{\label{asep LDHD p=0.5} Color online.
    Finite-size scaling plot of $\rho_n(t)$ for $p=0.5$ parallel-update TASEP 
    in the high-density and low-density phases, 
    for $L=10^3, 5\times10^3,10^4$ and a variety of $\alpha,\beta$.
    The choices of $\alpha,\beta$ shown correspond to four distinct 
    values of $v_c$ providing strong evidence for the conjecture 
    \eqref{ASEP tau int conjecture}.}
\end{figure}

Figs.~\ref{asep LDHD p=0.25} and~\ref{asep LDHD p=0.5} show
finite-size scaling plots of $\rho_n(t)$  for $p=0.25,0.5$, assuming
the ansatz given by \eqref{rho conjecture} and \eqref{ASEP tau int conjecture}.
There is again excellent data collapse, however we note
that there is some noticeable curvature near the edge of the support,
so that the sharp cusp present in the $p=0$ case  becomes smoothed out
somewhat for $p>0$.  As discussed in section 
\ref{infinite support section}, this does not affect the use of 
\eqref{tau int with discrete corrections} for setting Monte Carlo error bars, 
but it would be interesting from a theoretical perspective to better 
understand how this curvature depends on the model parameters $p,\alpha,\beta$
and $L$  
(as well as $v_{\max}$; c.f. the discussion in section~\ref{NaSch}).
We remark that many other quantities (including the fundamental diagram)
have cusps at $p=0$ which are smoothed out for $p>0$. 

\section{Nagel-Schreckenberg model}
\label{NaSch}
An important generalization of the TASEP is the Nagel-Schreckenberg
model~\cite{NagelSchreckenberg92}, in which each particle (vehicle)
can move up to $v_{\max}\in\mathbb{N}$ sites per iteration.  Although
the precise form of the phase diagram depends on $v_{\max}$, the NaSch
model exhibits, in general, the same three qualitatively distinct
phases as the TASEP~\cite{BarlovicHuisingaSchadschneiderSchreckenberg02}.
We now briefly review the dynamical rules defining
the NaSch model.  Suppose at time $t\in\mathbb{N}$ a vehicle with
speed $v_t\in\{0,1,\ldots,v_{\max}\}$ is located on site $x_t$, and
has {\em headway} (number of empty sites to its right) equal to $h_t$.
Then the maximum speed this vehicle can safely achieve at the next
time step is taken to be $v_{\text{safe}}= \min(v_t + 1, v_{\max}, h_t)$, 
which allows for unit acceleration provided  the speed limit is
obeyed and crashes are avoided.  Provided $v_{\text{safe}} > 0$, a
random deceleration is then applied so that with probability $p$ the
new speed is $v_{t+1}=v_{\text{safe}} - 1$, otherwise 
$v_{t+1} = v_{\text{safe}}$. Finally, in the bulk of the system,  the vehicle
hops $v_{t+1}$ sites to its right, so that $x_{t+1}=x_t + v_{t+1}$. 
All vehicles in the bulk of the system are updated in this way in parallel.
The bulk dynamics clearly reduces to parallel-update TASEP when $v_{\max}=1$.

It remains to consider the boundary dynamics.  We again wish to apply
open boundary conditions, however choosing an appropriate
implementation of such boundary conditions for the NaSch model is
actually surprisingly subtle, and has been an active topic of research
over recent years 
\cite{CheybaniKerteszSchreckenberg01a,CheybaniKerteszSchreckenberg01b,%
  Huang01,BarlovicHuisingaSchadschneiderSchreckenberg02,JiaMa09,%
  NeumannWagner09}. In particular, it was argued
in~\cite{BarlovicHuisingaSchadschneiderSchreckenberg02} that in order
to observe the maximum-current phase when $v_{\max}>1$ one needs to
implement the  inflow of vehicles into the system in a rather careful
manner.

Since our interest in the present context is confined to the high and
low density phases however, we have chosen to implement the boundary
conditions in the following simple way.  We augment the system, which
has sites $1\le i \le L$, with two boundary sites; one at $i=0$ and
another at $i=L+1$.  With probability $\alpha$ a vehicle with speed
$v_{\max}$ is inserted on site $0$, and we immediately compute
$v_{\text{safe}}$ for this vehicle.  If $v_{\text{safe}}>0$ we move
the vehicle to site $v_{\text{safe}}$ otherwise we delete it.  The
output is performed similarly.  With probability $1-\beta$ we insert a
vehicle on site $L+1$, which then acts as a blockage to vehicles
exiting the system.  If the rightmost vehicle in the system has 
$x_t \ge L - v_{\max}$ we define its new speed to be $v_{\text{safe}}$ and
attempt to move the vehicle to site $x_{t+1} = x_t + v_{\text{safe}}$.
If $x_{t+1}>L$ the vehicle is removed from the system.  When
$v_{\max}=1$ the above prescription reduces to the boundary rules for
the simple TASEP described in section~\ref{introduction}.

\subsection{Simulations}
We now describe our simulations of the NaSch model as defined above.
To our knowledge, no rigorous results are known for $v_c$ when
$v_{\max}>1$. However, for the deterministic case ($p=0$) we expect that
\begin{equation}
  v_c =
  \begin{cases}
    v_{\max},           & \text{low density phase},  \\
    -1,                 & \text{high density phase}, \\
  \end{cases}
  \label{deterministic nasch vc conjecture}
\end{equation}
for any $v_{\max}$, since in the low-density phase
the deterministic movement of vehicles from left to right should control
the dynamics, while in the high-density phase we expect that it is the
movement of holes (traveling with speed 1) from right to left which is
important. More generally, we expect the form \eqref{TASEP collective velocity}
to remain valid, but with an unknown function $g$, 
that will in general depend on $v_{\max}$.

\begin{figure}[ht]
  \includegraphics[scale=0.6]{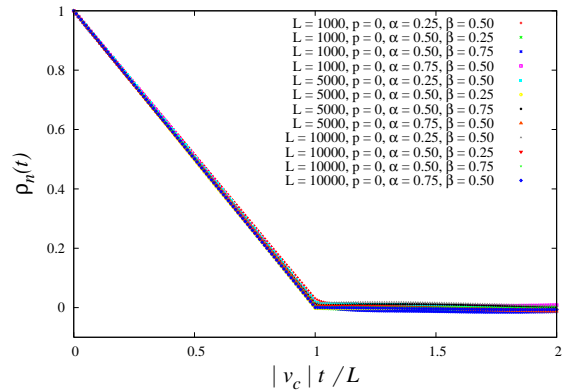}
  \caption{\label{nasch p=0 LDHD} Color online.
    Finite-size scaling plot of $\rho_n(t)$ for $p=0$ NaSch with $v_{\max}=3$ 
    in the high-density and low-density phases, 
    for a variety of choices of $\alpha,\beta$ and $L$.
    The exact value of $v_c$ is unknown in this case but here we chosen 
    $v_c$ according to \eqref{deterministic nasch vc conjecture}.}
\end{figure}

Fig.~\ref{nasch p=0 LDHD} presents a
finite-size scaling plot of $\rho_n(t)$ obtained by simulating the
NaSch model with $v_{\max}=3$ and $p=0$, with system sizes $L=10^3$,
$5\times 10^3$ and $10^4$ and a variety of values of $\alpha,\beta$
corresponding to both the low and high density phases.  The data
collapse is excellent, providing strong evidence for the ansatz
obtained from \eqref{rho conjecture}, \eqref{tau conjecture}, and
\eqref{deterministic nasch vc conjecture}.  As for the case of $p=0$
when $v_{\max}=1$ we note the sharpness of the cusp at $t=L/|v_c|$,
again suggesting that any corrections to the ansatz \eqref{rho conjecture}
are very small.
Each simulation performed consisted of $10^4 L/|v_c|$ iterations (with
$v_c$ given by \eqref{deterministic nasch vc conjecture}), with the
first $10^3 L/|v_c|$ time-steps discarded.
The above simulations were also performed for $v_{\max}=5$
with identical results.

\begin{figure}[ht] 
  \includegraphics[scale=0.6]{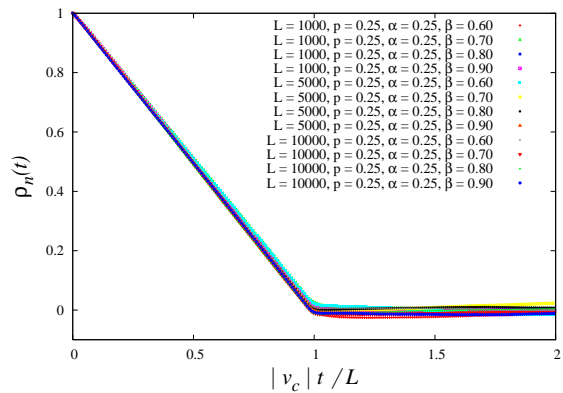}
  \caption{\label{nasch p=0.25 LDHD} Color online.
    Finite-size scaling plot of $\rho_n(t)$ for $p=0.25$ NaSch with 
    $v_{\max}=3$ in the high-density and low-density phases, 
    for a variety of choices of $\alpha,\beta$ and $L$.
    The exact value of $v_c$ is unknown in this case but here we 
    have set $v_c=2.65$.}
\end{figure}

Finally, we also considered the case of $v_{\max}=3$ with $p=0.25$.
For $v_{\max}>1$ and $p>0$ we are not aware of any exact predictions
for $v_c$, however it seems reasonable to conjecture that $v_c$ is
independent of $\beta$ ($\alpha$) in the low (high) density phase.  We
therefore simulated the NaSch model with $v_{\max}=3$, $p=0.25$ and
$\alpha=0.25$ at four different values of $\beta>\alpha$, which should
then correspond to a single value of $v_c$.  By considering a single
value of $v_c$ we can still use a finite-size scaling plot of
$\rho_n(t)$ to test the conjectures \eqref{rho conjecture} and
\eqref{tau conjecture}.  Fig.~\ref{nasch p=0.25 LDHD} provides strong
evidence to support their validity at $v_{\max}>1$ and $p>0$.  By
varying the value of $|v_c|$ used to produce the scaling plot of
$\rho_n(t)$ so that the support edge lay at $|v_c|t/L=1$ we obtained
$v_c\approx 2.65$.  We remark that, assuming the validity of 
\eqref{rho conjecture} and \eqref{tau conjecture}, this method can be used as a
way to obtain approximate values of $|v_c|$ when $v_{\max}>1$ and
$p>0$.

\section{Discussion}
\label{discussion}
We have studied the NaSch model in the low and high density phases via
Monte Carlo simulation, and found that to a very good approximation
the autocorrelation function for the system density behaves as
$1-|v_c\,t|/L$ with a finite support $[-L/|v_c|,L/|v_c|]$, where $v_c$
is the collective velocity.  For the case of $v_{\max}=1$ an exact
theoretical result is known for $v_c$ for all $p\in[0,1]$.  When
$v_{\max}>1$ no rigorous results for $v_c$ are known, however we
conjecture that the when $p=0$ we simply have $v_c=v_{\max}$ in the
low-density phase and $v_c=-1$ in the high-density phase.  This result
agrees with the exact result in the special case of $v_{\max}=1$ and
with numerical simulations for $v_{\max}=3,5$.
It seems reasonable to expect that it
is valid for all $v_{\max}$ for the deterministic NaSch model.

\begin{acknowledgments}
This research was supported by the Australian Research Council.
ZZ acknowledges support from the NSFC under Grant No. 10975127 and the NSF of Anhui under Grant No. 090416224.
TMG would like to thank Alan Sokal for some useful comments.
\end{acknowledgments} 

\bibliographystyle{apsrev}

\end{document}